\documentclass[superscriptaddress,amsmath,amssymb,aps,notitlepage,twocolumn,floatfix]{revtex4-2}
\usepackage{cmbright}
\DeclareFontShape{OT1}{cmss}{m}{it}{<->ssub*cmss/m/sl}{}

\usepackage{graphicx}% Include figure files
\usepackage{verbatim}
\usepackage{tabularx}
\usepackage{dcolumn}% Align table columns on decimal point
\usepackage{bm}% bold math
\usepackage{pict2e}
\usepackage{nicefrac}
\usepackage{braket}

\usepackage[colorlinks=true, allcolors=blue,breaklinks=true]{hyperref}
\usepackage[dvipsnames]{xcolor}

\DeclareMathOperator{\re}{Re}
\DeclareMathOperator{\im}{Im}

\DeclareMathOperator{\indref}{ref}

\newcommand{\specificthanks}[1]{$+$}

\begin{document}

\title{Statistical learning of engineered topological phases \\
in the kagome superlattice of \texorpdfstring{{AV}$_3${Sb}$_5$}{Lg}}

\author{Thomas Mertz}
%\thanks{These two authors contributed equally}
\email{mertz@itp.uni-frankfurt.de}
\affiliation{Institut f\"ur Theoretische Physik, Goethe-Universit\"at, 60438 Frankfurt am Main, Germany}
\author{Paul Wunderlich}
%\thanks{These two authors contributed equally}
\email{wunderlich@itp.uni-frankfurt.de}
\affiliation{Institut f\"ur Theoretische Physik, Goethe-Universit\"at, 60438 Frankfurt am Main, Germany}
\author{Shinibali Bhattacharyya}
\affiliation{Institut f\"ur Theoretische Physik, Goethe-Universit\"at, 60438 Frankfurt am Main, Germany}
\author{ Francesco Ferrari}
\affiliation{Institut f\"ur Theoretische Physik, Goethe-Universit\"at, 60438 Frankfurt am Main, Germany}
\author{Roser Valent\'i}
\email{valenti@itp.uni-frankfurt.de}
\affiliation{Institut f\"ur Theoretische Physik, Goethe-Universit\"at, 60438 Frankfurt am Main, Germany}

\date{\today}

\begin{abstract}
Recent experimental findings have reported the presence of unconventional charge orders in the enlarged ($2\times 2$) unit-cell of kagome metals AV$_3$Sb$_5$ (A=K,Rb,Cs) and hinted towards specific topological signatures. 
Motivated by these discoveries, we investigate the types of topological phases that can be realized in such kagome superlattices. In this context, we employ a recently introduced  statistical method capable of constructing topological models for any generic lattice.
By analyzing large data sets
generated from symmetry-guided distributions of randomized tight-binding parameters, and labeled with the corresponding topological index, we extract physically meaningful information. We illustrate the possible real-space manifestations of charge and bond modulations
and associated flux patterns for different topological classes,
and discuss their relation to present theoretical predictions and experimental signatures for the AV$_3$Sb$_5$ family. Simultaneously, we predict new higher-order topological phases that may be realized by appropriately manipulating the currently known systems. 
\end{abstract}

\maketitle

\vspace{0.2cm}{\bf\large Introduction}

The recent surge of interest in kagome materials, often discussed in the context of frustrated magnetism and spin liquid phases~\citep{balents2010spin,savary2016quantum,mendels2007quantum,han2012fractionalized,review_norman,review_broholm,jeschke2013first}, has been boosted by the discovery of the kagome metals AV$_3$Sb$_5$ (A=K, Rb, Cs) undergoing successive charge density wave (CDW)  and superconducting transitions upon lowering temperature~\citep{Ortiz_new_kagome_materials,SC_temp_KVSb,Ortiz_PRL_2020,SC_temp_RbVSb}.
The presence of flat bands, Dirac points, and van-Hove singularities in the electronic band structure of the ideal kagome lattice provides a playground for exotic topological properties and a variety of phases, ranging from superconductivity to charge, orbital momentum, and spin density waves~\citep{syozi_kagome,kiesel_kagome,mazin_2014,guterding_2016,Balents_Landau_2021,Thomale_GL_2021,jiang2021kagome}.
Density functional theory calculations for  AV$_3$Sb$_5$ have categorized the normal-state of this family as a $\mathbb{Z}_2$ topological metal with multiple protected Dirac crossings~\citep{Ortiz_PRL_2020} and renormalization group analyses have proposed the occurrence of various complex CDW and charge bond order (CBO) phases~\citep{Balents_Landau_2021,Nandkishore_CDW_2021,Slager_arxiv_2021,Thomale_GL_2021}. 
Interestingly, reports of giant extrinsic anomalous Hall effect suggest nontrivial band topology 
in the absence of long range magnetic order~\citep{Tyrel_AHE_2020}, possibly driven by a CDW order with orbital currents, and 
 high-resolution 
STM (scanning tunneling microscopy) measurements point to an unconventional intrinsic chiral charge~\citep{Neupert_Thomale_Nature_2021,Hasan_STMChiral_2021,Wang_STMChiral_2021,zhao2021cascade} 
 consistent with a doubling of the unit-cell ($2\times2$ superlattice)~\citep{Wilson_XRD_2021}. 
These observations imply the relevance of the ubiquitous chiral charge order present in the Haldane model~\citep{Haldane1988}, and the possibility of higher-order topological insulators, an avenue that demands further exploration. Although a thorough understanding of the various electronic orders warrants detailed microscopic investigations~\citep{christensen2021theory, Thomale_GL_2021, Nandkishore_CDW_2021, Balents_Landau_2021}, we use the available plethora of experimental observations as our motivation to learn about the possible topological phases that can manifest within the electronic parameter space of the $2\times2$ kagome superlattice.

Another rapidly developing field of research is the application of machine learning to tackle physical problems~\citep{carrasquilla2020}, from variational representation of wave functions~\citep{Carleo_2017}, to the detection of phase transitions~\citep{Carrasquilla2017_NatPhys, Zhang2019_EQM}. 
Due to the absence of a local order parameter, topological phase transitions are generally more difficult to capture than symmetry-breaking phase transitions, although some progress has been achieved~\citep{Melko_PRB2018, Rodriguez2019_topology,greplova2020unsupervised}. 
Additionally, an immediate physical interpretation of the results can turn out to be a complicated task in unbiased machine learning approaches.
Yet, in a recent study~\citep{Mertz_2021} we proposed a statistical learning of topological models on a honeycomb lattice and showed that machine-assisted unbiased learning can differentiate between the electronic parameters that are most significant for the manifestation of the well-known topological Haldane phase in the underlying lattice structure. 
Making use of a generalization of this method~\citep{Mertz_1x1kagome}, in this work we extract topological information for the generic $2\times2$ kagome superlattice, as “learned” from the statistics of data-sets of randomized tight-binding parameters, constrained only by specific crystal symmetries. The variations of the tight-binding parameters can be interpreted as modified hoppings arising from changes in lattice parameters, atomic mass, effects of strain, pressure, spin-orbit coupling, etc. 
We find topologically trivial Star-of-David (SoD)-like CBO phases and non-trivial chiral flux phases. Our results are compatible with present theoretical predictions and experimental observations for the AV$_3$Sb$_5$ family. Additionally, we predict new higher-order topological phases which might be realized in future experimental work through appropriate manipulation of the kagome lattice.

\vspace{0.4cm}{\bf\large Model}

\begin{figure}[t!]
    \centering
    \includegraphics[width=1\columnwidth]{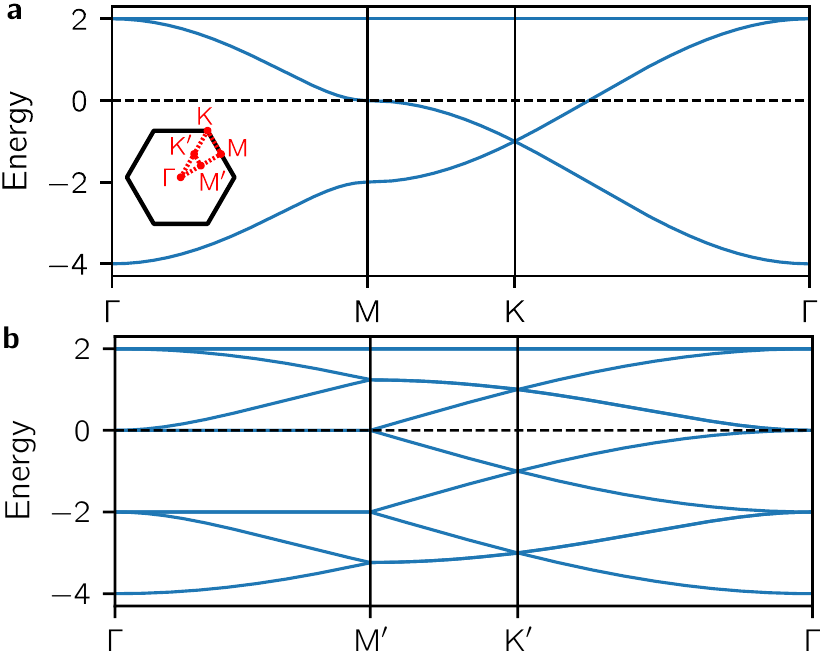}
	\caption{Metallic band structure along the high-symmetry path \textbf{a.}~$ \Gamma$-M-K-$\Gamma$ for the kagome lattice, and \textbf{b.}~$\Gamma$-M$^\prime$-K$^\prime$-$\Gamma$ for the $2\times 2$ kagome superlattice, for the tight-binding model with uniform nearest-neighbor hopping ($t_{i,j}=-1$) and $\epsilon_i=0$. The dashed horizontal line indicates the Fermi energy at $\nicefrac{5}{12}$ filling. The inset of \textbf{a} shows the hexagonal Brillouin zone of the kagome lattice (in black) together with the location of the high-symmetry points and the chosen path (in red).
	}
    \label{fig:bands}
\end{figure}

 We consider a generic tight-binding Hamiltonian on the kagome lattice which can be taken, in a first approximation, as a minimal model to describe the low-energy electronic properties of the vanadium $3d$ bands in AV$_3$Sb$_5$~\citep{Ortiz_new_kagome_materials, gu2021gapless}
\begin{align}\label{eq:tb_hamiltonian}
    H = \sum_i \epsilon_i^{\phantom{\dagger}} c_i^\dagger c_i^{\phantom{\dagger}} + \sum_{\left<i,j\right>} t_{i,j}^{\phantom{\dagger}} c_i^\dagger c_j^{\phantom{\dagger}} 
\end{align}
Here $\langle i,j \rangle$ runs over nearest-neighbor sites and $c_i^{\phantom{\dagger}}$ ($c_i^\dagger$) annihilates (creates) an electron at site $i$.  $\epsilon_i$ denote onsite potentials, while $t_{i,j}$ are hopping integrals between sites $i$ and $j$.
In the simple case of uniform hopping, i.e., $t_{i,j} = -1$, and zero onsite potentials, $\epsilon_i=0$, the band structure of the system, shown in Fig.~\ref{fig:bands}a, is characterized by a flat band at high energy and two lower-lying dispersive bands that touch each other in a Dirac point at the corners of the BZ ($K$ points), and exhibit van Hove singularities at the $M$ points. 

\begin{figure}[t!]
    \centering
    \includegraphics[scale=1]{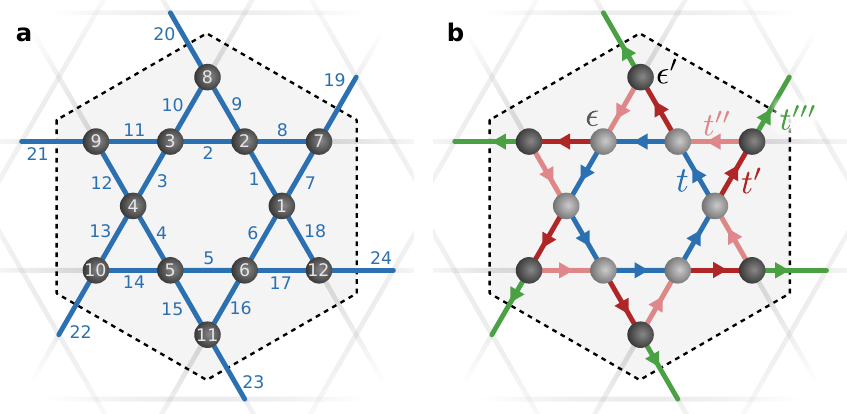}
     \caption{\textbf{a.}~Symmetric $2\times2$ kagome supercell. The unit cell, delimited by black dashed lines,
     contains 12 distinct sites. The 24 independent nearest-neighbor links are colored in blue. Sites and links are labeled counter-clockwise. \textbf{b.}~Schematic representation of the hoppings and onsite potentials in the $2\times2$ unit cell with $C_6$ symmetry. Spheres of different colors represent different onsite terms. Symmetries dictate equality of hoppings, as indicated by the colors of the bonds. The direction of the arrows signifies hopping from site $j$ to site $i$ (i.e., $c_i^\dagger c_j^{\phantom{\dagger}}$), and has been chosen arbitrarily, however, adhering to the constraints imposed by the $C_6$ symmetry.
     }
    \label{fig:2x2}
\end{figure}

Several works on AV$_3$Sb$_5$~\citep{Balents_Landau_2021,Thomale_GL_2021, Nandkishore_CDW_2021} have suggested that CDW instabilities at the van Hove fillings may cause a translational symmetry breaking of the perfect kagome lattice, leading to a lower periodicity described by a $2\times2$ supercell, analogous to that observed in STM experiments~\citep{Neupert_Thomale_Nature_2021,Hasan_STMChiral_2021,Wang_STMChiral_2021,zhao2021cascade}. 
For this reason, we focus at a filling $\nicefrac{5}{12}$ with the Fermi energy lying at the higher van Hove singularity (see Fig.~\ref{fig:bands}a). We assume our Hamiltonian to be periodic over the 
$2\times2$ enlarged unit cell 
illustrated in Fig.~\ref{fig:2x2}a which contains 12 sites (corresponding to the onsite terms $\epsilon_{i=1,\dots, 12}^{\phantom{\dagger} }$) arranged in a SoD pattern, and retains all the symmetries of the point group of the kagome lattice ($D_6$). 
As a consequence of the superlattice periodicity, we are left with 24 independent nearest-neighbor hoppings, which we label as $t_{s=1,\dots,24}$ as depicted by the blue-colored links in Fig.~\ref{fig:2x2}a. The hopping parameters can be categorized in three distinct classes: the hoppings in the inner hexagon ($t_{s \leq 6}$), the hoppings on the spikes of the SoD ($t_{7 \leq s \leq 18}$), and the hoppings connecting sites belonging to different unit cells ($t_{19 \leq s \leq 24}$). The various hoppings of each class can be mapped into each other by point group symmetry operations.
In the uniform case ($t_{s}=-1, \ \forall s$) with zero onsite potential, one obtains 12 bands as shown in Fig.~\ref{fig:bands}b for the BZ corresponding to the $2\times2$ superlattice. These bands can be unfolded to the 3 bands corresponding to the elementary $1\times1$ unit cell. By tuning the different hopping parameters, it is possible to open a topologically non-trivial gap at $\nicefrac{5}{12}$ filling We remark that the tight-binding model of Eq.~\eqref{eq:tb_hamiltonian} is not strictly bound to the description of the vanadium $3d$ bands of AV$_3$Sb$_5$ compounds, but retains a certain level of generality and could be applicable to other kagome systems at the van Hove filling. To investigate the possible topological phases of the $2\times 2$ kagome superlattice, we employ the statistical approach described in the Methods section.

\begin{figure*}[t!]
    \includegraphics[width=0.98\linewidth]{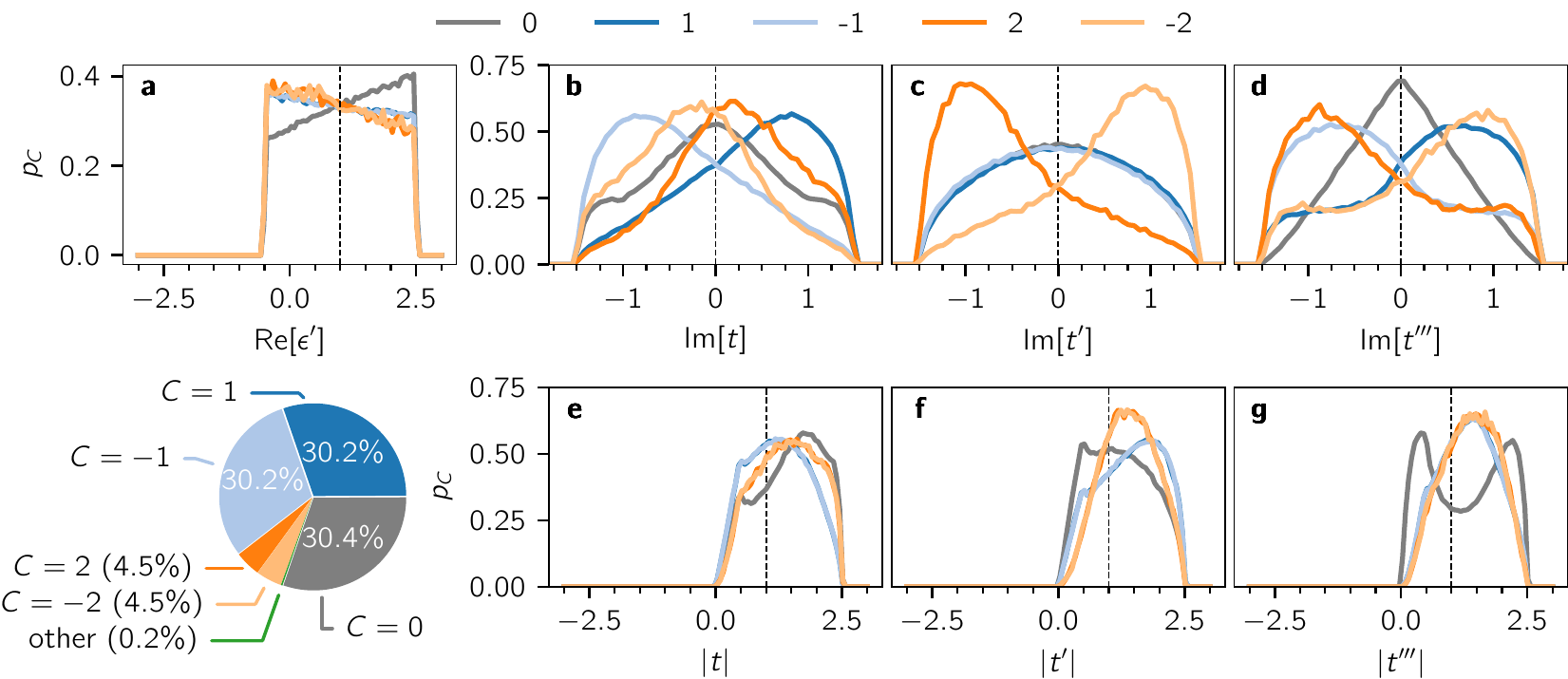}
    \caption{Distribution of Chern numbers (pie chart) and (\textbf{a}-\textbf{g}) marginal PDFs for the most descriptive features of insulators within the $2\times 2$ kagome superlattice with $C_6$ symmetry. The pie chart shows the percentages of trivial $(C=0)$ and topological $(C\neq0)$ samples obtained out of all insulating samples. For the PDF plots corresponding to different feature components, grey lines denote trivial phases $(C=0)$ and colored lines denote non-trivial $(C\neq0)$ phases as indicated by the legend on top. Dashed vertical lines indicate the reference point.}
    \label{fig:marginals}
\end{figure*}

\vspace{0.2cm}{\bf\large Results}

 A completely unbiased analysis of the full nearest-neighbor $2\times2$ kagome superlattice involves sampling of 11 onsite and 24 hopping parameters (one onsite term is kept fixed to set the global energy scale). To improve tractability, we scale down the number of independent features by enforcing specific symmetry operations on the feature space, for instance, the point group $C_6$, which is a subgroup of the kagome point group $D_6$, lacking reflection symmetries. This specific choice is necessary in order to construct non-trivial tight-binding models, since the Chern number---which we set as our
 topological index---is odd under the effect of reflections, and thus a reflection-invariant Hamiltonian could only have $C=0$.

Under $C_6$ symmetry, the onsite terms and hoppings are reduced to a set of six unique features as illustrated in Fig.~\ref{fig:2x2}b. These are real $\epsilon_{i\leq6} = \epsilon$ and $\epsilon_{i\geq7} = \epsilon^\prime$, and complex $t_{s\leq6} = t$, $t_{s} = t^{\prime}$ for $s \in \{7,9,11,13,15,17\}$, $t_s = t^{\prime\prime}$ for $s \in \{8,10,12,14,16,18\}$, and $t_{s\geq19} = t^{\prime\prime\prime}$. 
We choose $\mathbf{x^{\indref}} = { (\epsilon, \epsilon^\prime, t, t^{\prime}, t^{\prime\prime}, t^{\prime\prime\prime}) = }(0, 1, -1, -1, -1, -1)$ as reference point. Our choice for the sampling radius ensures that $\epsilon=0$ for all samples. We generate a data set of $n_\text{s} = 2\times 10^6$ samples, 67\% (33\%) of which are insulators (metals) \footnote{The classification of the samples in metals and insulators is performed numerically by computing the energy bands on a grid of $82\times 82$ $\textbf{k}$-points, and checking whether the indirect gap at $\nicefrac{5}{12}$ filling is smaller or larger than an energy threshold (chosen to be 0.01$ |t_{\indref}|$ here).}. As shown by the pie chart in Fig.~\ref{fig:marginals}, 
69.6\% of the insulating samples are topologically non-trivial. Among the topological insulators, the largest fraction (60.4\%) has $C=\pm 1$, while the second largest fraction (9\%) has $C = \pm 2$. In the following, we discuss the properties of these topological insulators in detail.

In the analysis of the hopping parameters we will focus on the marginal probability distribution functions (PDFs)
(see Eq.~\ref{Eq_PDF} of Methods) of the
onsite energy $p_C(\re{[\epsilon^\prime]})$ (Fig.~\ref{fig:marginals}a), imaginary parts of the hoppings $p_C(\im[t_s])$ 
[Figs.~\ref{fig:marginals}(b-d)], which determine the hopping direction, and PDFs of their moduli $p_C(|t_s|)$
[Figs.~\ref{fig:marginals}(e-g)], which describe the overall hopping strength. These features are those which provide most of the information about the topological character of the samples.
Due to the inherent symmetry of the kagome lattice, the PDFs for $t^\prime$ and $t^{\prime \prime}$ show the same behavior. Hence, the distribution for only one of these hoppings, i.e., $t^\prime$, is shown.
All PDFs are provided in Supplementary Fig.~1.
First, we analyze the PDF of the onsite term $\epsilon' = \epsilon_{i\geq7}$ (Fig.~\ref{fig:marginals}a) that distinguishes between trivial { (grey line) } and topological phases { (colored lines)}. For the trivial phase, $\epsilon'$ tends to be larger than zero, i.e., the outer ring of the spikes tend to be ``heavier'' compared to the inner hexagon. By contrast, in the topological phase $\epsilon'$ tends to have smaller values. This behavior is well known from the Haldane model~\citep{Haldane_QHE_wo_LL} and reflects the fact that large $|\epsilon^\prime|$ eventually turns the system into a trivial insulator.

\begin{figure}[t!]
    \centering
    \includegraphics[scale=1]{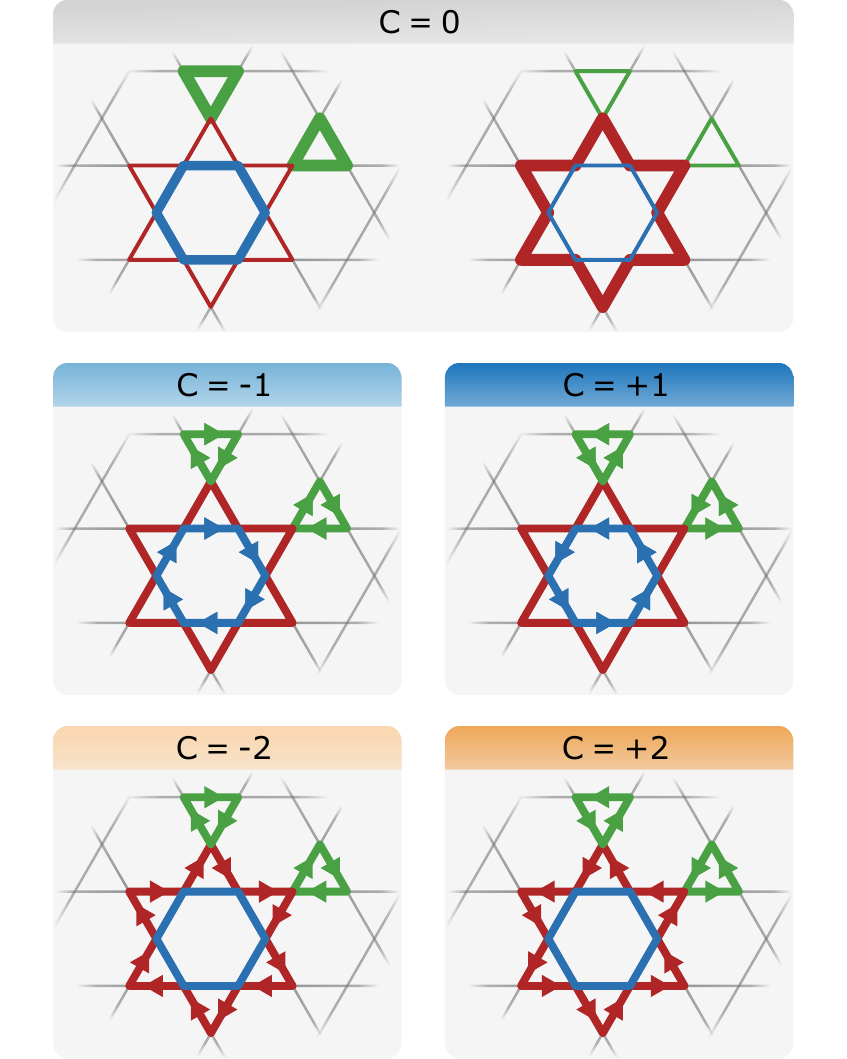}
    \caption{Overview of the characteristics of possible insulating phases of the $2\times2$ kagome superlattice with $C_6$ symmetry that is extracted from an unbiased data set. The hopping parameter $t$ is colored in blue, $t^\prime$ and $t^{\prime\prime}$ in red and $t^{\prime\prime\prime}$ in green. $t^{\prime\prime\prime}$ connects different unit cells and are arranged in triangular manner. Arrows illustrate the hopping flux direction and the line thickness indicates the relative magnitude of $|t|$, $|t^\prime| = |t^{\prime\prime}|$ and $|t^{\prime\prime\prime}|$. For $C=1$ and $C=2$ the hopping flux direction is mirrored with respect to $C=-1$ and $C=-2$, respectively. The handedness of the complex hopping patterns testifies the chiral nature of the topological phases.
    }
    \label{fig:phases_2symm}
\end{figure}

{\bf Trivial phases (C=0).} In the trivial phase $(C=0)$ we observe that the PDFs  $p_0(\im[t])$, $p_0(\im[t^\prime])$ and $p_0(\im[t^{\prime\prime\prime}])$ [Figs.~\ref{fig:marginals}(b-d)] show a maximum at zero and perfectly symmetric behavior around it. Hence, no particular hopping direction is preferred. 
The moduli $|t|$ and $|t^\prime|$ [Figs.~\ref{fig:marginals}(e-f)] tend to be slightly larger than $1$ (the reference value), and their PDFs do not show any significant structure. On the other hand,  $p_0(|t^{\prime\prime\prime}|)$ (Fig.~\ref{fig:marginals}g) possesses two local maxima of similar magnitude. By restricting the data set to the samples with $|t^{\prime\prime\prime}| < 1.25$ and $|t^{\prime\prime\prime}| > 1.25$ [corresponding to the approximate midpoint between the two local maxima of $p_0(|t^{\prime\prime\prime}|)$, see Supplementary Fig.~2], we identify two distinct dominant configurations with $C=0$, which are illustrated schematically in Fig.~\ref{fig:phases_2symm} (top row). 
One of them shows strong $|t|$ and $|t^{\prime\prime\prime}|$, and weaker $|t^\prime|$, consistent with an inverse Star of David (iSoD)-like CBO pattern (Fig.~\ref{fig:phases_2symm}, top row, left panel). The fraction of samples with this configuration amounts to 48\% of the trivial cases. The remaining 52\% of the trivial samples show an opposite pattern similar to the SoD-like CBO pattern, with larger $|t^\prime|$, and smaller $|t|$ and $|t^{\prime\prime\prime}|$ (Fig.~\ref{fig:phases_2symm}, top row, right panel). Such CBO patterns have also been predicted by phenomenological analyses of possible electronic instabilities at the van Hove filling~\citep{kiesel_kagome,Balents_Landau_2021, Nandkishore_CDW_2021, Jiangping_arxiv2021}, and STM experiments have hinted towards the presence of chiral charge order patterns~\citep{Neupert_Thomale_Nature_2021} in KV$_3$Sb$_5$ with an iSoD-like CBO as observed for the trivial phase. 
The $C=0$ phase of our analysis, however, is not chiral, since the \emph{real} hoppings of the tight-binding model fulfill all mirror symmetries of the kagome superlattice. On the other hand, the topological phases discussed in the remainder of the paper possess a chiral character due to complex hoppings, which induce non-trivial fluxes with a specific handedness (see Fig.~\ref{fig:phases_2symm}).

{\bf Topological phase (C=$\boldsymbol{\pm}$1).}
In contrast to the trivial phase, the hoppings in the topological phases display a preference for certain winding directions, as can be seen from the distributions $p_{C\neq 0}(\im[t])$, $p_{C\neq 0}(\im[t^\prime])$ and $p_{C\neq 0}(\im[t^{\prime\prime\prime}])$ in Figs.~\ref{fig:marginals}(b-d). 
Phases with positive Chern number can be distinguished from the corresponding phases with negative Chern number by the sign of the imaginary parts of the hoppings, since their respective PDFs are mirror images of each other. 
The PDFs of the moduli [Figs.~\ref{fig:marginals}(e-g)], instead, are equal for phases with positive and negative Chern number, and hence, do not distinguish between them. 
By restricting the data set to specific feature values, we analyze the most likely configurations for the respective Chern numbers.
\begin{figure}[b!]
    \centering
    \includegraphics[scale=1]{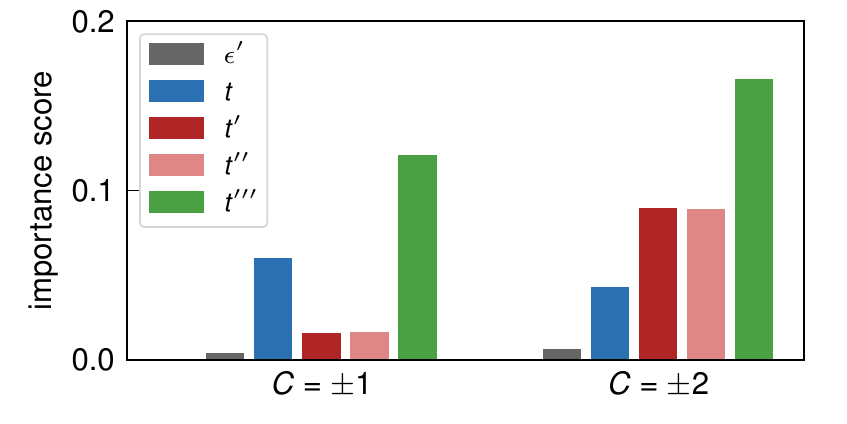}
     \caption{Importance score defined by the Bhattacharyya distance $D_B(p_{C}(x_i), p_{0}(x_i))$ (Eq.~\ref{eq:importance}) for phases with $C=\pm 1$ (left) and $C=\pm 2$ (right) of the $C_6$ symmetric model for the $2\times 2$ kagome superlattice. The distinct features $x_i$ are $\epsilon$, $\epsilon^\prime$, $t$, $t^\prime$, $t^{\prime\prime}$ and $t^{\prime\prime\prime}$ as defined in the main text.}
    \label{fig:importance}
\end{figure}

We evaluate the \emph{importance score} $D_B(p_C(x_i), p_{0}(x_i))$ (see Eq.~\ref{eq:importance} of the Methods Section) to identify the most descriptive features $x_i$ that distinguish the non-trivial phases from the trivial one. The results are shown in Fig.~\ref{fig:importance}. The importance score of $\epsilon$ is trivially zero since it is always kept at a constant value of $\epsilon = 0$. Due to the large overlap of $p_0(\epsilon^\prime)$ and $p_{C\neq0}(\epsilon^\prime)$, the importance of $\epsilon^\prime$ is rather low. However, as described earlier, the PDFs show clear peaks revealing $\epsilon^\prime$ as distinguishing parameter between topological and trivial phases. Next, we infer from Fig.~\ref{fig:importance} that $t^\prime$ and $t^{\prime\prime}$ have the same importance, since their PDFs show the same behavior. $t$ and $t^{\prime\prime\prime}$ have higher importance than $t^\prime$ (and $t^{\prime\prime}$) for differentiating $C=\pm 1$ phases from the $C=0$ phase, while $t^\prime$ (and $t^{\prime\prime}$) and $t^{\prime\prime\prime}$ are more important than $t$ for differentiating $C=\pm 2$ phases from the $C=0$ phase. This importance with respect to the differentiation between the Chern classes is reflected further in the PDFs of these features in a qualitative manner, as discussed in the following.

For $C=\pm 1$ phases, we find that the moduli $|t|$, $|t^{\prime\prime}|$ and $|t^{\prime\prime\prime}|$ behave similarly
[Figs.~\ref{fig:marginals}(e-g)], and we infer that the relative bond strengths may not be a strong distinguishing feature for the topological phases. This is depicted in Fig.~\ref{fig:phases_2symm} (middle row) by equal thickness of the blue, red and green bonds for $C=\pm 1$ phases.
On the other hand, we gain crucial insight from the imaginary parts of hoppings in this phase [Fig.~\ref{fig:marginals}(b-d)]. 
For $C=1$, both $\im[t]$ and $\im[t^{\prime\prime\prime}]$ tend to be larger than zero, which corresponds to a counter-clockwise winding of the hoppings of the inner hexagon and a clockwise winding of the hoppings forming the outer triangles (connecting different $2\times 2$ cells), as schematically illustrated by arrows in Fig.~\ref{fig:phases_2symm} (middle row).
The sign of $\im[t^\prime]$ and $\im[t^{\prime\prime}]$, instead, does not discriminate between $C=1$ and $C=-1$ due to missing contrast between the corresponding probability distributions. Hence, the orientations of $t'$ and $t''$ bonds are not shown in Fig.~\ref{fig:phases_2symm} for $C=\pm 1$. 
We note that a large fraction of $C=1$ topological insulators (49\%) shows this configuration, while the remaining samples are distributed incoherently.

Our characterization of the $C=\pm 1$ phase shares similarities with the ``chiral flux phase" (CFP) proposed in Ref.~\citep{Jiangping_chiralflux_2021} as a minimal model for the time-reversal symmetry breaking which is observed in muon spin relaxation experiments in KV$_3$Sb$_5$~\citep{mielke2021timereversal} and CsV$_3$Sb$_5$~\citep{yu2021evidence}, and for the giant anomalous Hall effect measurements~\citep{Tyrel_AHE_2020} in KV$_3$Sb$_5$. 
The CFP phase, which represents a possible electronic instability of the kagome metal at the van Hove filling~\citep{Balents_Landau_2021}, is described by a $C_6$-symmetric tight-binding model, which breaks time-reversal, but is invariant under the simultaneous action of time reversal and lattice reflections~\citep{Jiangping_chiralflux_2021}, analogous to the Haldane model on the honeycomb lattice~\citep{Haldane1988}. As opposed to the CFP phase of Ref.~\citep{Jiangping_chiralflux_2021}, our results for the $C=\pm 1$ phase suggest that the imaginary parts of  $t^\prime$ and $t^{\prime\prime}$ hoppings may not play a relevant role in the topological character of this phase.

{\bf Topological phase (C=$\boldsymbol{\pm}$2).}
In the $C= \pm2$ phases, the moduli of all features behave similarly to the $C = \pm 1$ phases. On the other hand, the sign of $\im[t]$ does not discriminate between $C=2$ and $C=-2$ due to low contrast between the PDFs $p_2(\im[t])$ and $p_{-2}(\im[t])$ [Fig.~\ref{fig:marginals}~(b)].
Phases with positive and negative Chern number are differentiated by the signs of $\im[t^\prime]$, $\im[t^{\prime\prime}]$ and $\im[t^{\prime\prime\prime}]$, which leads to their relatively higher importance score. For $C=2$, the hoppings along the outer spikes of the SoD ($t^\prime$ and $t^{\prime \prime}$) show clockwise winding, while the hoppings in the outer triangles ($t^{\prime \prime\prime}$) show counter-clockwise winding, as illustrated in Fig.~\ref{fig:phases_2symm} (bottom row). The largest coherent group of samples of topological insulators with Chern number $C=2$ (56\%) shows this particular configuration. 

{\bf Further analysis.}
Motivated by recent experimental results  detecting signatures of rotational symmetry breaking in the electronic properties of some {AV$_3$Sb$_5$} materials~\citep{li2021rotation,jiang2021kagome}, we investigate the fate of the topological phases of Fig.~\ref{fig:phases_2symm} when the symmetry of the tight-binding model is reduced from $C_6$ to $C_2$. We repeat our statistical analysis by forcing the Hamiltonian on the $2\times 2$ superlattice to be invariant only under rotations of 180${}^\circ$, thus increasing the feature space of the model to 18 distinct parameters, i.e., 6 onsite potentials and 12 hoppings. We start from a reference point (as explained
in the Methods Section) with uniform hoppings ($t=-1$) and zero onsite potentials. Among our samples, only a fraction of 13\% represent topological insulators (vs.~46\% in the analysis with $C_6$ symmetry), most of them possessing $C=\pm 1$ Chern number (98.6\%). While a smaller fraction of topological insulators can be expected as a consequence of the enlargement of the feature space, the strong reduction of $C=\pm 2$ samples (1.4\% of the total number of topological insulators) may suggest that this phase is rather fragile to rotational symmetry breaking. In contrast, the $C=\pm 1$ insulating phase is considerably less affected, and thus seemingly more stable. 

It is worth emphasizing that, although our analysis has been performed on a model of spinless electrons which explicitly breaks time-reversal symmetry (due to the presence of complex hoppings), our results can provide direct information about what one shall expect for a time-reversal invariant Hamiltonian of spinful electrons. Indeed, in analogy to the generalization of the Haldane model to the Kane-Mele model~\citep{KaneMele2005}, we took two copies of our topological tight-binding Hamiltonians of Fig.~\ref{fig:phases_2symm} to construct a  time-reversal invariant model for spinful electrons. The samples with odd Chern number in the spinless case, i.e., those belonging to the $C=\pm 1$ phase, yield a non-trivial $\mathbb{Z}_2$ invariant in the case of spinful electrons, which is characteristic of quantum spin Hall phases~\citep{KaneMele2005}.

\vspace{0.2cm}{\bf\large Summary and outlook}

By employing machine-assisted unbiased statistical learning constrained only by specific crystal symmetries, we extract meaningful topological information concerning the $2\times2$ kagome superlattice.
The highlights of our procedure are three-fold: first, one is able to tune through a large parameter space to find non-trivial topology in the kagome superlattice, second, specific crystal symmetries can constrain these parameters resulting in certain flux patterns concomitant with CBO/CDW orders, and third, one retains high levels of physical interpretability of the results.
For the kagome superlattice with $C_6$ symmetry, we infer possible SoD/iSoD-like CBO patterns and topologically non-trivial flux patterns from the large data sets of randomized hopping parameters. 
Our findings for the trivial and topological phases share similarities with recent experimental observations and theoretical predictions for the intensely discussed AV$_3$Sb$_5$ kagome materials.
Moreover, we infer that additional topological phases with higher Chern index $(C=\pm2)$ might exist which have neither been observed nor predicted yet.
Furthermore, by reducing the crystal symmetry to $C_2$, whose signatures were found in AV$_3$Sb$_5$ in recent experiments, 
we examined the stability of topological phases.
While $C=\pm1$ appears to be stable, the discovered $C=\pm2$ phases seem to be rather fragile. We also extended our analysis to spinful Hamiltonians, which show quantum spin Hall states. Our results provide a repository of knowledge that can guide future engineering endeavors to build kagome materials
(or modify existing ones) with a desirable topological phase. 
In this regard, a foreseeable extension of the present work consists of pursuing a material-specific analysis, searching for topological phases in the feature space of a tight-binding model obtained by \textit{ab initio} calculations for a specific target material. In the case of AV$_3$Sb$_5$ compounds, this may involve a multi-orbital description, featuring additional vanadium $d$-orbitals and antimony $p$-orbitals~\cite{gu2021gapless}, and the inclusion of spin-orbit coupling effects. Furthermore, the investigation of a layered superlattice geometry, such as the $2\times2\times4$ structure recently observed in Raman spectroscopy~\cite{wu2022charge} and x-ray diffraction~\cite{xiao2022coexistence} experiments in CsV$_3$Sb$_5$, represents a viable future direction. In both cases, the addition of physically motivated ingredients in the tight-binding Hamiltonian could lead to an improved understanding of the actual physical origin of the chiral topological phases.

\vspace{0.2cm}{\bf\large Methods}

 As introduced in Refs.~\citep{Mertz_2021, Mertz_1x1kagome}, our statistical approach uses random number generators to yield a data set of $n_s$ different tight-binding Hamiltonians (\emph{samples}) for a given lattice. Each sample is characterized by a vector of \emph{features}, \mbox{$\mathbf{x} = (x_1, \dots,x_{n_\text{f}})$}, grouping the $n_f$ distinct parameters of the model, and is classified by the \emph{label} $l$, which is a function of the features, i.e., $l = f(\mathbf{x})$. 
 For the current system, the onsite terms $\epsilon_i$ and the (complex) hopping parameters $t_i$ act as features. Samples are categorized into metals and insulators based on the presence of a finite band gap at the filling to be considered. After omitting metallic samples, the first Chern number $C$~\citep{Berry, Zee_non_abel_C, Fukui_chern_algo} is then chosen as the label for insulating samples. Hence, a feature vector for a sample is given by $\mathbf{x} = (\epsilon_1, \epsilon_2, \dots, t_1, t_2, \dots)$ with label $l = C[H(\mathbf{x})] \in \mathbb{Z}$.

Each sample in the data set is generated by randomly picking a value for each feature from a uniform probability distribution function (PDF). Specifically, for a given complex feature $x_i$, where $i$ indexes different features,
we sample the uniform PDF restricted to a sphere in the complex plane centered at a given reference point $x^{\indref}_i$, with radius $\alpha |x^{\indref}_i|$, where $\alpha \in \mathbb{R}$. Throughout this work, we choose $\alpha = 1.5$. This choice for the sampling space ensures physically reasonable configurations, since extreme hopping values are excluded. For gaining maximum insight into the data, we can decompose the complex features $x_i$ into real features, namely the real part $\re[x_i]$, the imaginary part $\im [x_i]$, the magnitude $|x_i|$, and the phase $\varphi[x_i]=\arg[x_i]$.

To understand which features play a major role in determining the topological properties of the model, we calculate the PDFs $p_l(x_i)$ of each feature $x_i$ for each label $l$. This is achieved by integrating out all other features $x_j\neq x_i$ from the bare distributions of the topological class $\rho_l(\mathbf{x})$
\begin{align}
    p_l(x_i) = \int \dots \int \rho_l(\mathbf{x}) \prod_{j\neq i} \text{d} x_j .
    \label{Eq_PDF}
\end{align}
For a given feature $x_i$, the comparison of the PDFs for different labels $l$, i.e., for different Chern numbers, provides information on the importance of $x_i$ for the topological properties of the tight-binding model. To quantify the difference between two PDFs, we make use of the Bhattacharyya distance~\citep{Bhattacharyya_distance}, defined for a complex feature as 
\begin{align}
    D_B(p,q) = - \log\left[ \int \sqrt{p(x_i) q(x_i)} \text{d} x_i \right] .
    \label{eq:importance}
\end{align}
Here, $p(x_i)$ and $q(x_i)$ are generic PDFs and $D_B(p,q)$ is always larger than zero unless $p=q$. 

The measure represented by $D_B$ acts as an indicator of the descriptiveness of features $x_i$ through $D_B(p_l(x_i), p_0(x_i))$, which quantifies the difference of the respective PDFs for Chern labels $l\neq 0$ w.r.t.~the trivial case. Larger values contribute most to the topological character. 
Based on this, one can simplify the investigation of the feature space by focusing only on the most descriptive features with high \emph{importance score}, which amounts to a dimensionality reduction. A complementary strategy makes use of symmetries that are either based on observed behavior of the PDFs or physical motivation.

The combined approach can generally take several iterations of re-sampling and analyzing the obtained data sets.
The interplay of different features can be assessed by computing statistical correlations among them. A straight-forward estimator of linear correlations is provided by the Pearson correlation coefficient~\citep{Pearson1895}.
A complementary way to investigate correlations is to restrict the data set to samples where certain features have specific values, e.g., $\im[x_i]<0$, and afterwards investigating the PDFs of the restricted data set, as done here. 

Summarizing, this approach tackles an $n_f$-dimensional phase space by sampling hopping parameters and computing the Chern number of the resulting Hamiltonians. From the average properties of the distributions of the different Chern numbers, we are able to reconstruct \emph{a posteriori} an effective description of the topological phases and their properties. 
This method not only yields information on the symmetry of the topological phases, but also provides crucial insights on which hoppings play a relevant role in determining the topological character. For example, the statistical analysis of 
the ${C=\pm 1}$ phases identified in our work indicates that the imaginary parts of $t'$ and $t''$ hoppings are not important to determine the topological character of the state, as discussed in the main text.

\vspace{0.2cm}
{\bf DATA AVAILABILITY}\\
The datasets generated and/or analysed during the current study are available from the corresponding authors upon reasonable request.

\vspace{0.2cm}
{\bf CODE AVAILABILITY}\\
The calculation codes used in this paper are available from the corresponding authors upon reasonable request.

\vspace{0.2cm}
{\bf\large Acknowledgments}\\
We thank R. Thomale for discussions. We acknowledge support from the Deutsche Forschungsgemeinschaft (DFG, German Research Foundation) through TRR 288-422213477 (Project B05) (PW,SB,RV) and through FOR 5249-449872909 (Project P4) (TM,RV). 
FF acknowledges support from the Alexander von Humboldt Foundation through a postdoctoral Humboldt fellowship.

\vspace{0.2cm}
{\bf\large Author contributions}\\  TM and PW performed the calculations and contributed equally to the work. SB and FF contributed to the analysis of the statistical data and the implementation of symmetries. RV supervised the project. All authors made contributions to the
development of the approach and wrote the paper.

\vspace{0.2cm}
{\bf\large Competing interests}\\
The authors declare no competing interests.

\vspace{0.2cm}
{\bf\large Additional information}\\
{\bf Supplementary information} The online version contains supplementary material available at ...

\bibliography{references}

\end{document}